\documentclass[aps,twocolumn,prl,superscriptaddress,amsmath,showpacs,tightenlines,nofootinbib]{revtex4-2}
\usepackage{epsfig,graphicx,times}
\usepackage{amstext}
\usepackage{amsmath}            
\usepackage{amssymb}            
\usepackage{graphicx}           
\usepackage{latexsym}
\usepackage{bm}
\usepackage{color}
\usepackage[pdfpagemode=UseNone,pdfstartview=FitH,colorlinks=true,linkcolor=blue,urlcolor=blue,anchorcolor=blue,citecolor=blue]{hyperref}
\makeatletter

\@ifundefined{textcolor}{}
{\definecolor{BLACK}{gray}{0}
 \definecolor{WHITE}{gray}{1}
 \definecolor{RED}{rgb}{1,0,0}
 \definecolor{GREEN}{rgb}{0,1,0}
 \definecolor{BLUE}{rgb}{0,0,1}
 \definecolor{CYAN}{cmyk}{1,0,0,0}
 \definecolor{MAGENTA}{cmyk}{0,1,0,0}
 \definecolor{YELLOW}{cmyk}{0,0,1,0}
}

\newcommand{\m}{\mathrm}

\makeatother
\begin{document}

\title{Dissipative acousto-mechanical parametric interface between high-overtone acoustics \\and flexural phonons}

\author{Xun Ji}
\thanks{These authors contributed equally to this work.}
\affiliation{Beijing Key Laboratory of Fault-Tolerant Quantum Computing, Beijing Academy of Quantum Information Sciences, Beijing 100193, China}
\affiliation{Beijing National Laboratory for Condensed Matter Physics, Institute of Physics, Chinese Academy of Sciences, Beijing 100190, China}
\affiliation{University of Chinese Academy of Sciences, Beijing 100049, China}

\author{Huanying Sun}
\thanks{These authors contributed equally to this work.}
\affiliation{Beijing Key Laboratory of Fault-Tolerant Quantum Computing, Beijing Academy of Quantum Information Sciences, Beijing 100193, China}
 
\author{Longhao Wu}
\affiliation{Department of Applied Physics, Aalto University, FI-00076 Aalto, Finland}

\author{Qichun Liu}
\affiliation{Beijing Key Laboratory of Fault-Tolerant Quantum Computing, Beijing Academy of Quantum Information Sciences, Beijing 100193, China}

\author{Yulong Liu}
\email{liuyl@baqis.ac.cn}
\affiliation{Beijing Key Laboratory of Fault-Tolerant Quantum Computing, Beijing Academy of Quantum Information Sciences, Beijing 100193, China}

\author{Mika A. Sillanpää}
\affiliation{Department of Applied Physics, Aalto University, FI-00076 Aalto, Finland}

\author{Tiefu Li}
\email{litf@tsinghua@edu.cn}
\affiliation{School of Integrated Circuits, BNRist and Frontier Science Center for Quantum Information, Tsinghua University, Beijing 100084, China}
\affiliation{Beijing Key Laboratory of Fault-Tolerant Quantum Computing, Beijing Academy of Quantum Information Sciences, Beijing 100193, China}
\date{\today}	
\begin{abstract}
High-overtone bulk acoustic wave resonators (HBARs) promise advanced phononics, yet achieving nonlinearity remains challenging. We demonstrate a radiation-pressure-type parametric interaction between GHz HBARs and low-frequency flexural modes in a suspended silicon nitride membrane, where mechanical displacement modulates the external dissipation rate to enable dissipative acousto-mechanical coupling. Benefiting from the high quality factor, the system enters the resolved-sideband regime at room temperature, yielding acousto-mechanically induced transparency. We observe tunable Kerr nonlinearity and generate coherent HBAR frequency combs via two-tone driving. Notably, our dissipative coupling strength is 20 times larger than the dispersive coupling, the highest ratio among reported hybrid dissipative‑dispersive coupling systems, resulting in the experimental observation of amplification in the reflection spectra under red‑sideband driving. The ability to interface dense HBAR modes with a common mechanical resonator provides a scalable on-chip platform for multimode phononic information processing, with quantum phononics potentially achievable at sub‑Kelvin temperatures.
\end{abstract}

\maketitle

High-overtone bulk acoustic resonators (HBARs) serve as versatile platforms by supporting high-order harmonics within macroscopic substrates. Their capacity for exceptionally high frequency-quality factor product ($f\times Q$) at gigahertz (GHz) frequencies has spurred widespread adoption across advanced phononic applications~\cite{Review-HBAR-NRM-2021,zhao2023review,Parker2025acousticclock,PhysRevLett.127.071102}. Recently, HBARs have emerged as a cornerstone in quantum acoustodynamics, owing to their demonstrated capability for phononic quantum information processing~\cite{HBAR-NP-acoustictoolbox,HBAR-Quantum-acoustics-Science-2017,HBAR-multimode-PRApplied-2022,HBAR-acoustodynamics-NC-2020,HBAR-NP-acoustodynamics-2022}. In the microwave frequency band, the piezoelectric effect enables strong coupling between HBARs and superconducting resonators, with interaction strengths reaching the megahertz (MHz) regime~\cite{AFC-MF-PRL-2022,PhysRevLett.117.123603}. Leveraging this electric-dipole-like coupling, strong interactions with superconducting qubits at the single-phonon level have been realized~\cite{HBAR-mechanical-qubit-Science-2024,HBAR-catstates-Science-2023,HBAR-qubit-PRL-2023,HBAR-qubit-APL-2020,HBAR-coupling-APL-2023,HBAR-interface-PRL-2019,HBAR-qubit-PRB-2018}, enabling deterministic preparation of non-Gaussian phononic states~\cite{HBAR-Fockstate-PRL-2025,HBAR-quantumstates-Nature-2018}, among others. In the optical domain, parametric coupling between HBARs and optical cavities has been explored through photoelastic effect~\cite{HBAR-opto-SA-2019,HBAR-Optomechanical-NP-2025,PhysRevApplied.23.054024,Renninger2018,PhysRevResearch.5.043140}, with recent demonstrations showing that HBAR-microring integration facilitates electro-optic modulation~\cite{doi:10.1126/sciadv.aar4994,doi:10.1126/science.adg3812,PRXQuantum.1.020315,1gvf-w6lx}. These systems have demonstrated the microwave-to-optical transduction efficiency approaches 1\% and is fundamentally limited by the electro-optic coefficients~\cite{Han:21-Optica,10.1063/5.0021088,Barzanjeh2022optomech}.

\begin{figure*}[ht]  
    \includegraphics[width=0.96\linewidth]{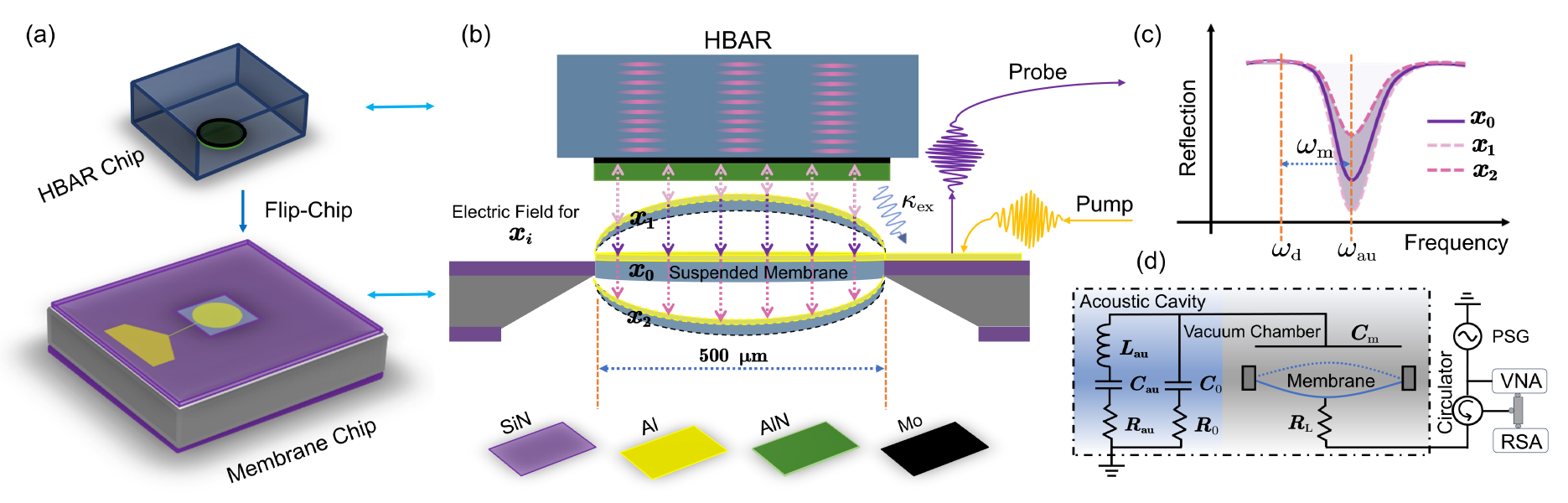}
    \caption{(a) Schematic of the dissipative acousto-mechanical interface (not to scale), featuring a high-overtone bulk acoustic resonator (HBAR) and a SiN membrane mechanical resonator integrated via flip-chip bonding. (b) Conceptual illustration of the coupling: the flexural displacement of the membrane modulates the HBAR's external dissipation rate $\kappa_{\text{ex}}$.  (c) Schematic of the probed reflection spectra $|S_{11}|$ for varying membrane displacements $x_i$ ($x_0 < x_1 < x_2$). The varying attenuation directly manifests the dissipative coupling. (d) Equivalent lumped-element circuit and experimental configuration. All measurements are conducted at room temperature in a high-vacuum chamber. PSG: multi-channel phase-coherent signal generator; VNA: vector network analyzer; RSA: real-time spectrum analyzer.}
    \label{F1}
\end{figure*}

Concurrently, two-dimensional (2D) membrane mechanical resonators, typically operating in the sub-MHz range, have emerged as pivotal components in quantum phononics~\cite{SiN-phonon-waveguide-Nature-2025,Degeneracy-breaking-NC-2025,SiN-optome-PRL-2024,SiN-memory-NPJ-2023,SiN-sensor-NC-2020,SiN-mircrowave-PRL-2021,Albert-bandgapSiN-NN-2017}. In the optical domain, low-loss silicon nitride (SiN) membranes are widely employed in high-$Q$ optical microcavities~\cite{SiN-RoomT-PRL-2016,SiN-lowloss-PRL-2014,SiN-OFC-MFC-NC-2023,SiN-Comb-NC-2023} to explore optomechanical modulation. Notably, by integrating bandgap-engineered SiN membranes within high-finesse Fabry–P\'{e}rot cavities, quantum control of mechanical motion has been achieved~\cite{SiN-roomT-Nature-2024,Albert-bandgapSiN-Nature-2018,PhysRevLett.126.113601}. Despite this progress, the substantial frequency mismatch between these low-frequency 2D flexural modes and superconducting qubits poses a significant barrier to direct interaction in the GHz band~\cite{SiN-electro-optic-Nature-2022, SiN-qubit-transduction-Nature-2020, AlN-membrane-NP-2018}. While flexural 2D membrane resonators offer superior parametric coupling to optical fields, and bulk acoustic waves (BAWs) provide ideal resonant frequencies for microwave qubit interfacing, a hybrid platform that bridges the advantages of both remains elusive~\cite{LiNbO3-M2O-Optica-2019,PhysRevA.103.053504,HBAR-OM-coupling-NPhotonics-2016,Holzgrafe:20Cavity,Electro-optic-Optica-2023,HBAR-M2O-PRA-2021,HBAR-optical-NC-2020}. This gap stems from the formidable challenge of engineering efficient parametric interactions between phononic modes across such disparate frequency regimes.
\par
In this work, we demonstrate a hybrid acousto-mechanical interface that enables parametric coupling between HBARs and low-frequency SiN membrane resonators.   Our design features a readout electrode patterned directly on the suspended membrane, which translates mechanical displacement into modulation of the HBAR’s external dissipation rate, $\kappa_{\text{ex}}$, thereby realizing a dissipative optomechanical-like coupling~\cite{PhysRevLett.102.207209,PhysRevLett.125.233601,dissipative-coupling-NJP-2013,PhysRevLett.105.056801,Kazouini2026,Barnard2019,dissipative-dispersive-coupling-NC-2019,Dissipative-dispersive-coupling-PRA-2020,Dissipative-coupling-NC-2024}.  This approach overcomes the long-standing challenge of achieving sizable dissipative coupling in integrated electromechanical systems. Moreover, it achieves operation in the resolved-sideband regime at room temperature, unlike conventional superconducting electromechanical systems that require cryogenic conditions. Experimentally, we observe acousto-mechanically induced transparency (AMIT) under low-power pumping. With high-power red-sideband pumping, we demonstrate amplification in the reflection spectrum. Furthermore, leveraging the intrinsic Kerr nonlinearity enables the generation of phase-coherent acoustic frequency combs (AFCs). This versatile on-chip architecture, capable of interfacing a vast manifold of HBAR modes with a common mechanical resonator, provides a scalable framework for exploring many-body phonons.
\par

\textit{Dissipative acousto-mechanical interface.}---As illustrated in Fig.~\ref{F1}(a), our device features a hybrid flip-chip architecture that integrates a high-frequency HBAR with a low-frequency flexural SiN membrane resonator.  To facilitate the acousto-mechanical interaction, the HBAR is accessed via a metallized SiN membrane that is held at a sub-micron distance in a flip-chip scheme. The HBAR is fabricated on a 250-$\mu$m-thick sapphire substrate, comprising a 60-nm-thick molybdenum (Mo) bottom electrode and a 1-$\mu$m-thick aluminium nitride (AlN) piezoelectric layer deposited via sequential sputtering. The mechanical resonator consists of a 50-nm-thick low-loss SiN membrane, grown via low-pressure chemical vapour deposition (LPCVD) on a silicon substrate. At the centre of the SiN chip, a 500~$\mu$m $\times$ 500~$\mu$m suspended window is coated with a 20-nm-thick aluminium (Al) layer. The metallization extends to an external coupler port of the HBAR, thus serving as a functional electrode that translates mechanical flexural motion into modulation of the HBAR's external coupling.

The side view of the flip-chip assembly is depicted in Fig.~\ref{F1}(b). The AlN layer serves as a high-efficiency electromechanical transducer, exciting a dense spectrum of longitudinal HBAR modes within the sapphire substrate. This substrate acts as a high-$Q$ acoustic Fabry–Pérot cavity, where the longitudinal modes are confined between the AlN/Mo interface and the sapphire/air boundary. The fundamental interaction mechanism hinges on the sensitive modulation of the acoustic cavity's external dissipation rate, $\kappa_{\text{ex}}$, by the out-of-plane displacement $x$ of the SiN membrane.  As shown in Fig.~\ref{F1}(c), each membrane displacement $x_i$ leads to a different external coupling strength, manifesting as a controllable contrast in the microwave reflection $|S_{11}|$ spectra.

To quantitatively describe the hybrid system, we employ the modified Butterworth-Van Dyke (MBVD) lumped-element model~\cite{HBAR-acoustodynamics-NC-2020}, as depicted in Fig.~\ref{F1}(d). Each overtone is represented by a motional branch consisting of a series $L_{\text{au}}$-$C_{\text{au}}$-$R_{\text{au}}$ circuit. The SiN membrane electrode and the Mo layer form a parallel-plate structure with a displacement-dependent capacitance $C_{\text{m}}(x)$, which shunts the HBAR motional branches. The total device impedance is given by:
\begin{equation}
Z = R_\mathrm{L} + \frac{1}{j\omega C_\mathrm{m}} + \left( Z_\mathrm{au} \parallel Z_0 \right).
\end{equation}
Here, $Z_{\text{au}} = R_{\text{au}} + j\omega L_{\text{au}} + 1/(j\omega C_{\text{au}})$ is the characteristic impedance of the acoustic mode, $Z_\textrm{0}=R_\textrm{0}+1/(j\omega C_\textrm{0})$ represents the dielectric loss of piezoelectric materials, and the external load $Z_\textrm{m}=R_\textrm{L}+1/(j\omega C_\textrm{m})$ is modulated by the mechanical mode.  In our electromechanical system, the membrane displacement modulates both the resonance frequency and the external dissipation rate of the HBAR. The respective single-photon coupling strengths are defined as $g_{\rm disp}=-\frac{\partial\omega}{\partial x} x_\mathrm{zpf}$ and 
$g_{\rm diss}=\frac{\partial\kappa_\m{ex}}{\partial x}x_\mathrm{zpf}$, where $x_{\text{zpf}}$ is the zero-point fluctuation of the SiN flexural mode. Under the resonance frequency approximation $\omega_{\rm au}=1/\sqrt{L_{\rm au}C_{\rm au}}$, analysis of the lumped-element network yields the ratio of the single-photon dispersive to dissipative coupling strengths $|\frac{g_{\rm disp}}{g_{\rm diss}}| = \frac{1}{2\beta(2-\eta)}$,
where $\beta = \omega_{\rm au} R_{\rm L} C_{\rm m}$ and $\eta = C_{\rm m}/(C_{\mathrm{0}}+C_{\rm m})$.
Indeed, the static capacitance $C_{\mathrm{0}}$ is much larger than the coupling capacitance $C_{\rm m}$ ($0<\eta \ll 1$).  Hence, when $\beta > 1/4$, dissipative coupling dominates, and consequently $|g_{\rm diss}| > |g_{\rm disp}|$.
In this regime, the dissipative coupling strength simplifies to $g_{\rm diss}= 2\omega_{\rm au}^2 R_{\rm L} \frac{C_{\rm m}}{C_{0}}\frac{dC_{\rm m}}{dx}x_{\rm zpf}$. As these relations indicate, increasing the coupling capacitor $C_{\rm m}$ directly enhances the relative strength of dissipative coupling (see Supplementary Material~\cite{SM}). The equivalent circuit model gives preliminary single-photon dissipative and dispersive couplings. The acousto-mechanical coupling model further clarifies their effective coupling contributions.
\par
We consider the interaction between this specific HBAR longitudinal mode centered at $\omega_{\text{au}}$ and the fundamental flexural mode of the SiN membrane at $\omega_{\text{m}}$. The single-mode description is legitimate, as the energy scales related to the mechanics and the interaction are much smaller than the free spectral range (FSR). To coherently control the interaction, the HBAR mode is driven by a strong microwave pump at frequency $\omega_{\text{d}}$, with a detuning $\Delta \equiv \omega_\m{d} - \omega_\m{au}$. A weak coherent probe tone with strength $\Omega_{\rm p}$ at frequency $\omega_p$ is simultaneously applied to the system. The dynamics are governed by the system-bath interaction Hamiltonian~\cite{PhysRevLett.102.207209,SM}. In the  rotating frame of the pump frequency $\omega_{\text{d}}$, the linearized equations of motion are obtained as:
\begin{align}
\dot{\hat{d}} =& - (iG_{\rm disp} + G_{\rm diss})(\hat{c}+\hat{c}^\dagger)+(i\Delta- \frac{\kappa}{2})\hat{d}
\nonumber\\&- \sqrt{\kappa}\,\hat{a}_{\rm in}-i\Omega_{\rm p} e^{i(\omega_{\rm d}-\omega_{\rm p})t}, \nonumber\\
\dot{\hat{c}} =&- iG_{\rm disp}(\hat{d}+\hat{d}^\dagger) + G_{\rm diss}\bigl( \alpha^* \hat{\xi}_{\rm in} + \hat{d}^{\dagger} - \hat{d} - \alpha \hat{\xi}_{\rm in}^{\dagger} \bigr) \nonumber\\
&-(i\omega_{\rm m}+ \frac{\gamma_{\rm m}}{2})\hat{c}- \sqrt{\gamma_{\rm m}}\hat{c}_{\rm in}. \label{eq:EOM}
\end{align}
Here, $\hat{d}{(\hat{d}^\dagger)}$ and $\hat{c}{(\hat{c}^\dagger)}$ are the annihilation (creation) operators for the fluctuations of cavity mode $\hat a$ and mechanical mode $\hat b$, introduced via the standard linearization $\hat a=\bar a+\hat{d}$ and $\hat b=\bar b+\hat{c}$. The mechanical damping rate is $\gamma_{\rm m}$. The total damping rate of the HBAR mode is $\kappa = \kappa_{\rm in} + \kappa_{\rm ex}$, where  $\kappa_{\rm in}$ is the internal dissipation rate and $\kappa_{\rm ex} = \kappa_{\rm c} + g_{\rm diss}\langle \hat b+\hat b^\dagger\rangle$ is the external dissipation rate modified by external coupling. $\kappa_c$ denotes the external coupling strength in the absence of any pump injection. Vacuum input noise through the drive port is denoted $\hat{\xi}_{\rm in}$. 
The steady-state intracavity field  $\bar{a}$ is related to the classical input amplitude $\bar{a}_{\rm in}$ by 
$\bar{a} = \alpha \bar{a}_{\rm in}$, with the cavity response coefficient given by $\alpha \approx \sqrt{\kappa_{\rm c}}/(\kappa/2 - i\Delta)$.
The enhanced dispersive and dissipative coupling rates are $G_{\rm disp}=g_{\rm disp}\bar{a}$ and $G_{\rm diss}=g_{\rm diss}\bigl(-\frac{i\Delta}{2\kappa_{\rm c}}+\frac{\kappa_{\rm in}}{4\kappa_{\rm c}}+\frac{1}{4}\bigr)\bar{a}$, respectively. In the cavity equation of motion, the mechanical displacement $\hat{c}+\hat{c}^\dagger$ couples to the cavity fluctuation via the complex coefficient $iG_{\rm disp}+G_{\rm diss}$. In the mechanical equation, the dissipative coupling additionally contributes a direct backaction term. In the resolved-sideband regime, $G_{\rm diss}$ is enhanced by a factor of $\Delta/\kappa_{\rm c}\approx 8$ in our system. Together with the intrinsic condition $|g_{\rm disp}/g_{\rm diss}|<1$, this leads to $|G_{\rm disp}|\ll|G_{\rm diss}|$, the dominance of dissipative coupling.
\begin{figure}[t]
    \includegraphics[width=1\linewidth]{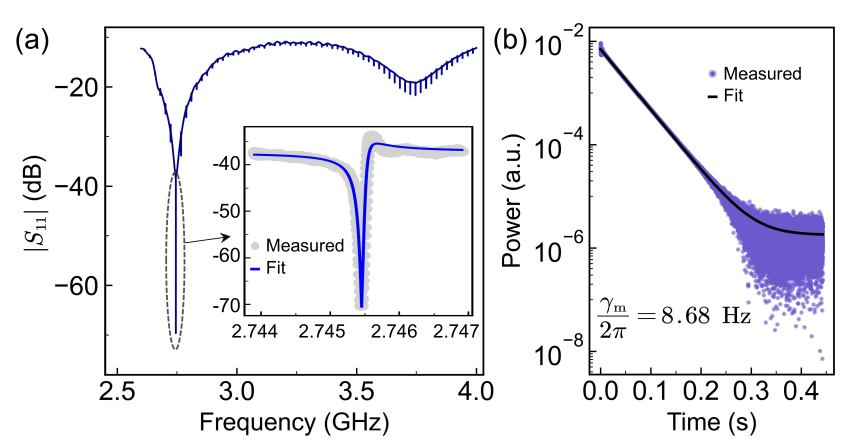}
    \caption{(a) Microwave reflection spectra $|S_{11}|$ of selected HBAR  modes. The inset shows a zoomed-in view of an acoustic mode along with its Lorentzian fit. (b) Ringdown measurement of the fundamental flexural mechanical mode of the SiN membrane.}
    \label{F2}
\end{figure}

In our experiments, all measurements were conducted in a high-vacuum chamber at room temperature. The microwave reflection spectrum, presented in Fig.~\ref{F2}(a), reveals a dense manifold of high-order longitudinal acoustic modes within the 2.5–4~GHz window. The measured FSR of $\mathcal{F}_0/2\pi \approx 21.37$~MHz matches the ballistic phonon model $\mathcal{F}_0 = v/2d$, where $v \approx 1.04 \times 10^4$ m/s is the longitudinal velocity in the sapphire substrate with thickness $d=250$ $\mu$m. In this high-$Q$ regime, the mode linewidths are governed by the interplay between intrinsic Akhiezer-type phonon scattering~\cite{HBAR-scattering-AIP-2024, Scatter-HBAR-PRB-2008} and surface boundary scattering, the latter becoming increasingly dominant as the phonon wavelength approaches the interface roughness at higher frequencies (see Supplementary Material~\cite{SM}). Specifically, we focus on a representative resonance at $\omega_{\text{au}}/2\pi=2.7454$~GHz, which exhibits a total quality factor of $Q \approx 1.89 \times 10^4$. This mode operates in the critical coupling regime ($\kappa_{\text{ex}} \approx \kappa_{\text{in}}$), with an external coupling rate $\kappa_{\text{ex}}/2\pi = 72.93$~kHz and an intrinsic dissipation rate $\kappa_{\text{in}}/2\pi = 72.5$~kHz.

The out-of-plane fundamental flexural mode of the SiN membrane with resonance frequency $\omega_\textrm{m}/2\pi=571.5$~kHz was characterized using VNA and RSA. By applying a red-detuned pump ($\Delta \approx - \omega_{\text{m}}$) and a resonant probe, we tracked the energy decay through the evolution of the in-phase and quadrature (IQ) components, as shown in Fig.~\ref{F2}(b). To minimise backaction-induced linewidth broadening and ensure the measurement reflects the intrinsic dissipation, these ringdowns were performed in the low-power limit. The time-domain ringdown of the quadrature components yields a relaxation time of $\tau = 36.7$~ms, from which the intrinsic mechanical damping rate is extracted as $\gamma_{\text{m}}/2\pi = 8.68 \pm 0.2$~Hz. Furthermore, the single-photon acousto-mechanical coupling strength is determined to be $g_0/2\pi=0.059$~Hz (see Supplementary Material~\cite{SM}). Next, using the same pump and probe setup, we also analysed the effective coupling of the system.

\begin{figure}[t]
    \includegraphics[width=1\linewidth]{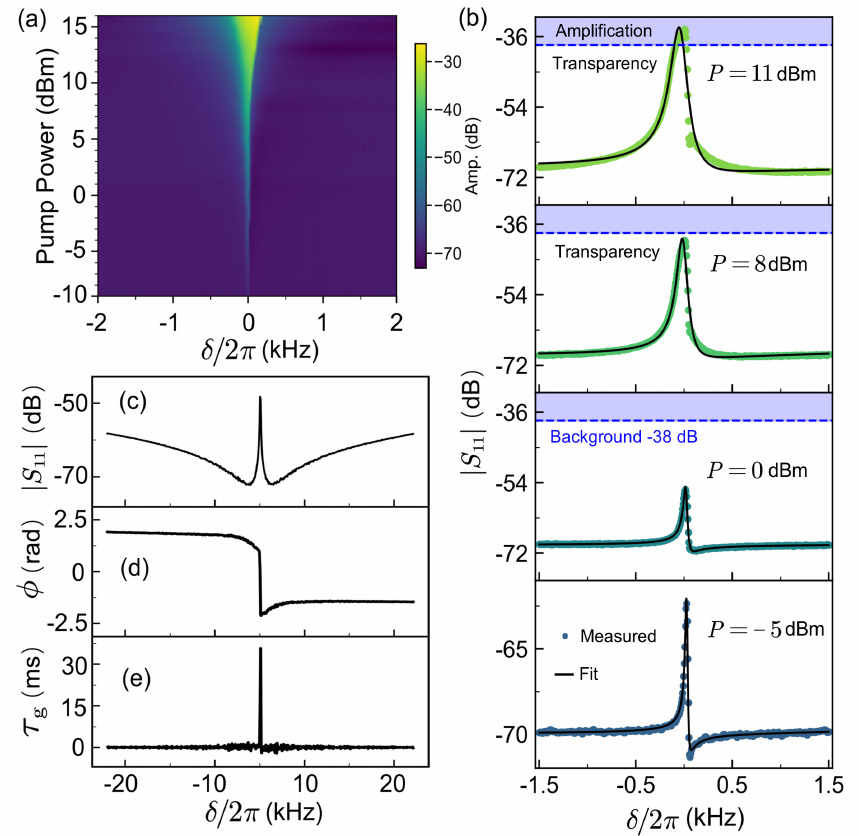}
    \caption{(a) Reflection spectra  $|S_{11}|$ of the AMIT window (color scale) as a function of red-detuned pump power. (b) Detailed view of the transition region from (a) for selected drive powers with the fitted AMIT (black lines). (c)–(e) Measured transparency window characteristics at a pump power of $3$~dBm. (c) the transmission amplitude of the AMIT peak, (d) the accompanying steep phase dispersion, and (e) the resulting group delay $\tau_{\text{g}}$ calculated from the phase derivative.}
    \label{F3}
\end{figure}

\textit{Acousto-mechanically induced transparency.}---In the resolved-sideband regime ($\omega_{\text{m}} > \kappa$), a strong red-detuned drive enables coherent control of the mechanical mode via the HBAR acoustic field. When a weak probe sweeps across the resonance and its detuning from the drive equals $\omega_{\text{m}}$, coherent oscillations generate Stokes and anti-Stokes sidebands. The destructive interference between the anti-Stokes field and the probe field manifests as a narrow transparency window in the acoustic reflection spectrum, termed acousto-mechanically induced transparency (AMIT), which signifies the suppression of probe absorption via a mechanically mediated destructive pathway. With the red-detuning anchored at $\Delta = -\omega_{\text{m}}$ and accounting for both acoustic and mechanical dissipation, the effective probe transmission spectrum $t_p(\delta)$ is given by:
\begin{align}
    t_p(\delta)
    =1-
    \frac{\kappa_{\rm ex}+\frac{|G_{\rm disp}G_{\rm diss}|\chi_{\text{m}}(\delta)\kappa_c}{2\Delta}\big(i+\frac{G_{\rm diss}}{2G_{\rm disp}\Delta\chi^*_{\rm au}(0)}\big)
    }
    {\frac{1}{\chi_{\rm au}(\delta)}+|G_{\rm disp}|^2\chi_{\text{m}}(\delta)[1+(\frac{G_{\rm diss}}{2G_{\rm disp}\Delta\chi_{\rm au}(0)})^2]}
        \label{eq:input_output}
\end{align}
where the relative probe detuning is $\delta = \omega_{\text{p}} - \omega_{\text{au}}$; $\chi_{\mathrm{au}}(\delta)$ and $\chi_{\text{m}}\left( \delta \right)$ represent the susceptibility of the acoustical cavity and mechanical mode, respectively. Correspondingly, the effective mechanical linewidth is modified by dissipative coupling compared to the purely dispersive case and can be expressed as $ \gamma_{\mathrm{eff}}\approx\gamma_{\rm m}+\gamma_{\rm b}$, where $\gamma_{\rm b}$ is the damping arising from the dynamical backaction exerted by the HBAR on the SiN membrane (see Supplementary Material~\cite{SM}).
\par  
The dependence of the AMIT window transmission on the pump power is shown in Fig.~\ref{F3}(a) and reveals a broadened linewidth and an increasing amplitude as the power increases. The zoomed-in spectra in Fig.~\ref{F3}(b) further reveal key signatures of dissipative coupling, including the asymmetry and a peak rising above the background baseline. Over the entire pump power range from –10 to 15~dBm, the baseline stays near –38~dB, increasing by only 0.2~dB. The solid lines represent fits based on the theoretical model. At pump power $P$ = -5 dBm, the fit yields $|G_{\rm disp}/G_{\rm diss}|\approx 0.2$ according to  Eq.~(\ref{eq:input_output}).  Since both $G_{\rm disp}$ and $G_{\rm diss}$ scale linearly with the intracavity field amplitude $\bar a$, the ratio $|g_{\rm disp}/g_{\rm diss}|$ remains approximately constant at 0.8 (see Supplementary Material~\cite{SM}). As the pump power increases, the system enters the nonlinear regime, introducing additional coupling contributions that cause
 $|g_{\rm disp}/g_{\rm diss}|$ to vary with the pump power. Meanwhile, dynamic backaction significantly reshapes the mechanical response.
 At $P$ = 11 dBm, the Fano-like fitting of AMIT yields $|G_{\rm disp}/G_{\rm diss}|\approx 0.05$, indicating that the effective dissipative coupling strength is 20 times larger than the dispersive coupling, the highest ratio among reported hybrid dissipative-dispersive systems.  
This confirms the dominance of dissipative coupling under strong pumping, which is enhanced by the factor $\Delta/\kappa_{\rm{c}}$ and $g_{\rm diss}$. 
A striking manifestation of this dominance is a reflection coefficient that exceeds the background level ($|S_{11}|>-38$ dB), a feature impossible with pure dispersive coupling. Finally, the single-photon coupling strength in the linear regime obtained by fitting 
$t_\text{p}$ in Fig.~\ref{F3} agrees in order of magnitude with the directly measured value $g_0/2\pi \sim 10^{-2}$~Hz (see Supplementary Material~\cite{SM}).

Then, a hallmark of AMIT is the associated slow‑acoustic effect, where the steep phase dispersion across the transparency window slows the propagation of an acoustic pulse. For probe pulses with bandwidths narrower than the effective linewidth $\gamma_{\text{eff}}$, the steep phase dispersion induces a significant group delay $\tau_{\text{g}} = -d\phi/d\delta$.
This is demonstrated in Fig.~\ref{F3}(c–e): panel (c) displays the transparency window at P=3 dBm, panel (d) shows the corresponding phase $\phi = \arg[t_\textrm{p}(\delta)]$ with its steep resonant slope, and panel (e) presents the derived group delay, reaching a maximum of about 35 ms.

\begin{figure*}[t]
    \includegraphics[width=0.97\linewidth]{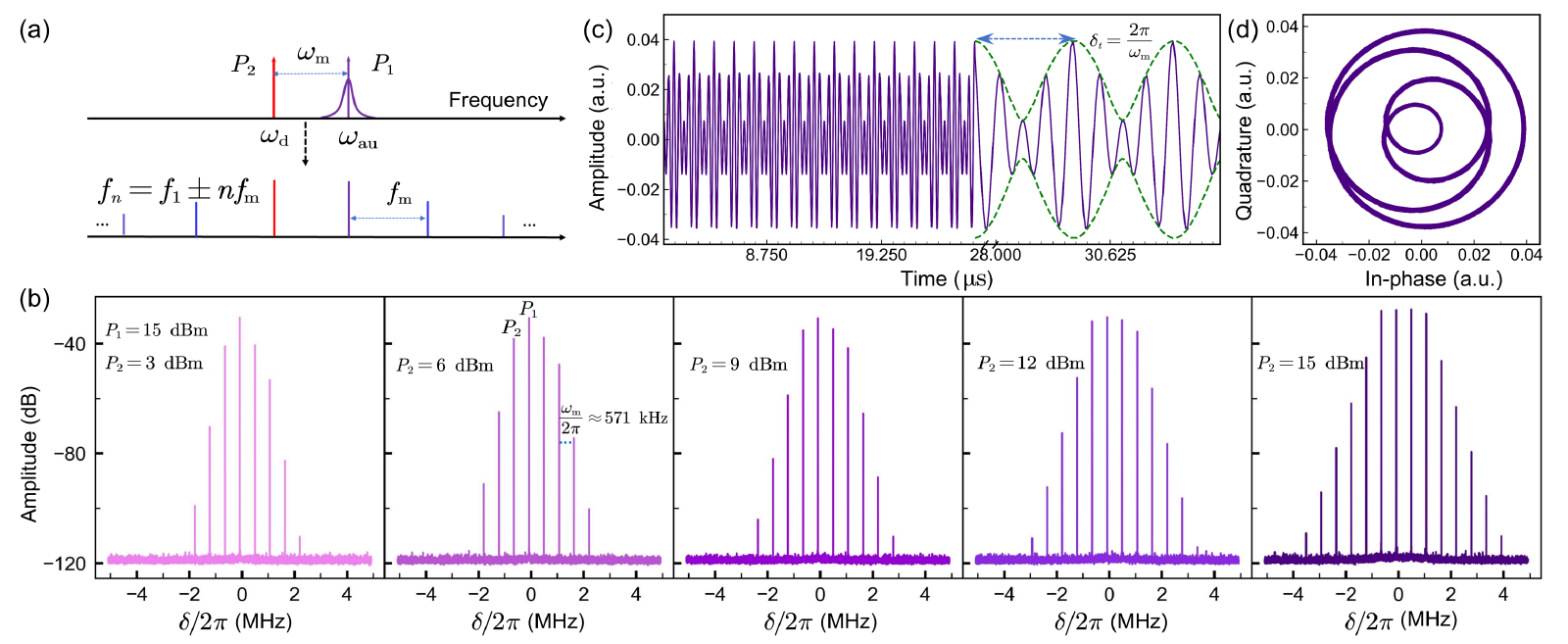}
    \caption{(a) Schematics of the two-tone pumping configuration, consisting of a resonant drive ($P_1$) and a red-sideband drive ($P_2$ at $\Delta \approx -\omega_{\text{m}}$).
    (b) Evolution of the AFC spectra measured by RSA under varying pump powers, with $P_1 = 15$ dBm fixed and $P_2$ varied. Controllable tooth counts and equidistant sidebands are observed. (c) Time-domain trace of the in-phase component, exhibiting clear periodic oscillations with $\delta_{t} = 1.75~\mathrm{\mu s}$. (d) Lissajous figure reconstructed from the IQ amplitudes, where the closed and stable orbits verify the long-range phase coherence of the generated frequency combs.}
\label{F4} 
\end{figure*} 

\textit{HBAR frequency combs via nonlinear acousto-mechanical coupling.}---The observed change in the ratio $g_{\rm disp}/g_{\rm diss}$ indicates that the system is being driven into a regime where the interplay between dispersive and dissipative couplings is strongly power-dependent. Beyond the linear regime, the radiation-pressure-type interaction gives rise to an effective Kerr nonlinearity. This is characterized by the Hamiltonian $\hat{\mathcal{H}}_{\text{int}}^{\text{non}} \approx \hbar \chi \hat{a}^\dagger\hat{a}^\dagger\hat{a}\hat{a}$, where the nonlinear coefficient $\chi \propto g_0^2$ originates from the second-order mechanical correction to the cavity field energy~\cite{MFC-Multimode-PRL-2022,Dongchunhua-OFC-PRL-2024}. In the strong-driving regime, the large-amplitude mechanical oscillation $X_{\text{m}}$ induces a periodic phase modulation on the HBAR cavity field, described by $a(t)=e^{-i \zeta \cos \left(\omega_{\text{m}} t\right)} \sum_n a_n e^{i n \omega_{\text{m}} t}$. Here, $\zeta = g_0 X_{\text{m}} / \omega_{\text{m}}$ represents the nonlinear modulation depth, and $a_n$ is the amplitude of the $n$-th sideband.

To harness the induced Kerr-type nonlinearity for spectral synthesis, we implement a two-tone driving scheme, as illustrated in Fig.~\ref{F4}(a).
By separately altering the resonant driving power ($P_1$) and the red sideband pump power ($P_2$), we achieve deterministic control over the spectral span of the resulting acoustic frequency comb (AFC) [Fig.~\ref{F4}(b)]. In the strong drive power regime ($15$~dBm), a robust AFC emerges with up to 14 well-resolved teeth, characterised by a uniform spacing precisely matched to the mechanical frequency $\omega_{\text{m}}$. The intensity envelope [Fig.~\ref{F4}(c)] and the corresponding Fourier transform [$P_2=15$~dBm in Fig.~\ref{F4}(b)] directly map the profile of the nonlinear modulation function. Crucially, the phase coherence of the generated AFC is rigorously verified through both temporal and phase-space analysis. Long-range phase coherence is corroborated by the periodic envelope of the in-phase component [Fig.~\ref{F4}(c)] and the stability of the Lissajous orbits reconstructed from the time-domain IQ data [Fig.~\ref{F4}(d)], where the period matches the mechanical cycle $\delta_{t} = 2\pi/\omega_{\text{m}} = 1.75~\mu\text{s}$, confirming that the modulation originates from the vibration of the fundamental mechanical mode of the SiN membrane. Operating at room temperature, the HBAR-based AFC provides a versatile platform for microwave synthesis, acoustic spectroscopy, and on-chip phononic signal processing, leveraging its GHz band and sub-MHz spacing.

\textit{Conclusion.}---We have proposed a hybrid acousto-mechanical platform that enables parametric coupling between high-overtone acoustics and flexural phonons, where mechanical displacement modulates the external dissipation rate of the acoustic cavity. We have achieved the highest reported ratio of dissipative and dispersive acousto-mechanical couplings ($|G_{\rm{diss}}/G_{\rm disp}|\approx 20$) to date. The high-quality factors of the HBAR modes enable operation within the resolved-sideband regime at room temperature, as demonstrated by the observation of both acousto-mechanically induced transparency (AMIT) and amplification (AMIA) even under red-sideband pumping. Furthermore, the Kerr nonlinearity of this interface enables the generation of phase-coherent acoustic frequency combs via a precise two-tone pumping scheme. 

Under ambient conditions, the device can serve as a room-temperature alternative to cryogenic superconducting electromechanical systems for precision metrology and sensing. Importantly, its intrinsic cryo-compatibility also renders it suitable for quantum-regime applications, where dissipative coupling could enable macroscopic quantum state preparation of flexural mechanical resonators, quantum state storage and transfer between distinct phononic modes. Furthermore, the ability to interface a dense manifold of HBAR modes with a mechanical resonator establishes robust and scalable architectures for high-capacity phononic information processing~\cite{Phononic-memory-NC-2017,Phononic-circuit-NPJ-2022,Scalable-phonoinc-NC-2025}. This versatile system opens new frontiers for exploring complex many-body dynamics, nonlinear phonon lasers, and topological transport in large-scale acoustic networks~\cite{Topologic-Networks-Phononic-PRL-2024,doi:10.1126/sciadv.adv9984, Phononic-circuit-NE-2025}. 

\textit{Acknowledgements.}---Y.~L. conceived the original idea, designed and fabricated the device, and coordinated the research. X.~J. performed the measurements with help of Q.~L. and drafted the manuscript with input from H.~S., L.~W., M.~A.~S., and Y.~L. All authors approved the final version of this work for publication. The authors acknowledge the support of the National Natural Science Foundation of China (No.~92365210, No.~12374325 and No.~12304387), Beijing Natural Science Foundation (Z240007), the National Key Research and Development Program of China (Grant No. 2022YFA1405200), European Research Council (contract 101019712), Research Council of Finland (Finnish Quantum Flagship, No.~358877) and the Young Elite Scientists Sponsorship Program by CAST (Grant No.~2023QNRC001). This work is also supported by Beijing Municipal Science and Technology Commission (Grant No.~Z221100002722011).

\textit{Data availability.}--—The data are not publicly available. The data are available from the authors upon reasonable request.

\bibliography{reference}

\end{document}